\documentclass[11pt]{article} 

\pdfoutput=1 

\begin{document} 
\title{Large-Eddy Simulation of an over-expanded planar nozzle} 
\author{Britton J Olson
\\\vspace{6pt} Sanjiva K Lele 
\\\vspace{6pt} Department of Aeronautics and Astronautics 
\\ Stanford University, CA 94305, USA}

\maketitle 

\begin{abstract} 
This fluid dynamics video shows visualizations of a Large-Eddy Simulation (LES) of an over-expanded planar nozzle.  This configuration represents the experimental setup of Papamoschou et. al.~\cite{Papam:10,Papam:09,Papam:06} which found the position of the internal shock to be unstable.  Our LES calculations~\cite{Olson:11} confirm this instability and offer a vibrant and dynamic view of the underlying flow physics.  The interaction between shock and turbulent boundary layer is shown as is the subsequent separation region downstream.  Numerical Schlieren provide a glimpse of the low frequency shock motion and suggest potential mechanisms for the instability.  Key features include the asymmetry of the shock structure (with large and small lambda shocks), compression and expansion waves downstream of the shock and large scale flow reversal.  Full details of the experiment and the calculation can be found in the references.
\end{abstract}

\end{document}